\documentclass[pra,superscriptaddress,twocolumn]{revtex4}

\usepackage{enumerate}
\usepackage{amsfonts,amssymb,amsmath}
\usepackage[]{graphics,graphicx,epsfig}
\usepackage{amsthm}

\usepackage{graphicx}
\usepackage{dcolumn}
\usepackage{natbib}
\usepackage{color}
\usepackage{multirow}
\usepackage{ulem}

\begin{document}

%%%%%%%%%%%%%%%%%%%%%%%%%%%%%%%%%%%%%%%%%%%%%%%%%%%%%%%%%%%%%%%%%%%%%%%%%%%%%%%%%%%%%%%%%%%%%%%%%%%%%%%%%%%%%%%%

\title{A little bit of classical magic to achieve quantum speedup}

\author{Dagomir Kaszlikowski}
\email{phykd@nus.edu.sg}
\affiliation{Centre for Quantum Technologies,
National University of Singapore, 3 Science Drive 2, 117543 Singapore,
Singapore}
\affiliation{Department of Physics,
National University of Singapore, 3 Science Drive 2, 117543 Singapore,
Singapore}

\author{Pawe{\l} Kurzy{\'n}ski}
\email{pawel.kurzynski@amu.edu.pl}
\affiliation{Faculty of Physics, Adam Mickiewicz University, Umultowska 85, 61-614 Pozna\'n, Poland}
\affiliation{Centre for Quantum Technologies,
National University of Singapore, 3 Science Drive 2, 117543 Singapore,
Singapore}

\date{\today}

%%%%%%%%%%%%%%%%%%%%%%%%%%%%%%%%%%%%%%%%%%%%%%%%%%%%%%%%%%%%%%%%%%%%%%%%%%%%%%%%%%%%%%%%%%%%%%%%%%%%%%%%%%%%%%%%

\begin{abstract}
We introduce {\it nebit}, a classical bit with a signed probability distribution. We quantify its entropy using Szekely theorem on signed probability measures and show that classical stochastic dynamics supplemented with nebits can achieve or even exceed the speedup of Grover's quantum search algorithm. The proposed classical dynamics never reveals negativity of nebits and thus we do not need any operational interpretation of negative probabilities. We argue that nebits can be useful as a measure of non-classicality as well as a tool to find new quantum algorithms.  

\end{abstract}

\maketitle

%%%%%%%%%%%%%%%%%%%%%%%%%%%%%%%%%%%%%%%%%%%%%%%%%%%%%%%%%%%%%%%%%%%%%%%%%%%%%%%%%%%%%%%%%%%%%%%%%%%%%%%%%%%%%%%%

\section{Introduction}
We know that quantum computers can be much faster at certain tasks than classical Turing machines. Flagship examples are the Shor's factoring algorithm \cite{Shor}, the Grover's search algorithm \cite{Grover} and boson sampling \cite{bosonsampling}. What we still, at least to some degree, do not know is why it so. It is clear that quantum superposition and quantum entanglement appear in quantum computations but how do they exactly contribute to quantum speed-up? Moreover, we do not have a measure of how non-classical any given quantum computation is. And perhaps most importantly, designing quantum algorithms is notoriously hard. Since the birth of the idea of quantum computation \cite{Benioff} we only know a handful of quantum codes in the quantum supremacy regime. The idea of classical dynamics with nebits is an attempt to address these increasingly important questions at the advent of rapidly developing quantum computers \cite{Google}.

This paper flirts with a non-orthodox or even iconoclastic ideas. Thus, to pre-empty an immediate dismissal, we note that these ideas were entertained long before by the other researchers (see for example \cite{reviewonnegative,Dirac,Feynman,Wigner,neg1,neg2,neg3,neg4,neg7,neg9,neg10}). For instance, R. Feynman studied negative probabilities in hope to resolve renormalisation issues of quantum electrodynamics \cite{Feynman}. E. Wigner was perhaps more successful, leaving behind a potent Wigner function formalism \cite{Wigner}, which is extensively used in quantum optics and other branches of quantum physics. Both Feynman and Wigner were extremely pragmatic in their approach to negative probabilities - as long as the final stage of calculations does not contain negative probabilities, everything is perfectly all right. In this sense, negative probabilities were for them just a computational tool and this is our modus operandi as well. This neatly removes any need to speculate about a meaning of negative probabilities and let us focus on more hardheaded tasks such as a deeper understanding of the advantages of quantum computer.  

Before we proceed with our arguments we would like to make another remark, following Wheeler who coined a phrase "great smokey dragon" as a poetic metaphor highlighting some counterintuitive aspects of quantum mechanics. It is well accepted by most physicists (if they are forced to make such a philosophical declaration) that quantum theory is an input-output black box process. You prepare a quantum state (input), evolve it (black box) and finally measure it (output). Wheeler's "great smokey dragon" symbolises strange internal workings of this quantum black box that do not have a classical equivalent. It is futile to look inside the box to see what happens because any observation disturbs and changes it unlike in classical theories where things exist objectively without a need for observation. Our approach parallels this quantum mechanical paradigm. We present a black box with internal classical dynamics supplemented with 'hidden' nebits whose output mimics the output of the Grover search algorithm on quantum computer. Nebits are never observed at the output, they are permanently locked inside of the black box. Again, in a truly classical box one could peek inside and see what is in there but we do not allow this as one of the rules of the game.

Finally, techniques developed in this paper, we hope, can be used as a tool to find new quantum algorithms. Such algorithms are notoriously difficult to come by. Mimicking quantum mechanics with classical stochastic dynamics supplemented with nebits might provide some intuitions or at least hints of how to translate it into purely quantum algorithms. 

\section{Nebit and its properties}
We first study some basic properties of a {\it nebit}. What follows is by no means a systematic study of this highly counterintuitive yet mathematically precise object. It's rather a bunch of loose observations that we need to explain the main ideas of this paper. 

Consider a binary system, a nebit, with states labelled as $0$ and $1$ in analogy to a classical bit. However, unlike for classical bits we allow these states to be taken with signed probabilities, i.e., probabilities that can be negative but still normalised to one: $p_0=1+\Delta$, $p_1=-\Delta$, where $\Delta \geq 0$. As we wrote in the Introduction, we do not attempt to give any operational interpretation to negative probabilities, which we take simply as objects obeying precise mathematical rules \cite{Szekely1}. Those rules, with a few irrelevant nuances, are that of Kolmogorovian probability theory.

\subsection{Entropy of nebit}

Szekely \cite{Szekely1} proved that any signed random variable $X$ can be always 'converted' to a non-negative random variable $Z$ (non-negative probability distribution) by adding to it a non-negative and statistically independent random variable $Y$. $Y$ can be seen as some kind of noise that masks the negativity of $X$.

Now, let us find $Y$ that converts our nebit to a proper bit (positive probability) whose Shannon entropy is zero. We have 
\begin{eqnarray}
p_z = \sum_{x+y=z}p_xp_y,
\end{eqnarray}   
where $y,z=0,1,2,\dots$ and $p_q$ ($q=X,Y,Z$) are probability distributions of the random variables $X,Y,Z$. To get $H(Z)=0$ we simply demand that $p_{z=0}=1$, which leads to the unique  solution for $Y$
\begin{equation}
	p_{y=k} = \frac{\Delta^k}{(1+\Delta)^{k+1}}.
\end{equation} 
As $X,Y$ are statistically independent, we have $H(Z)=H(X)+H(Y)$ and this gives us, for $H(Z)=0$,
\begin{eqnarray}
	H(X)&=&-H(Y)\nonumber\\
	&=&-p_{x=0}\log{p_{x=0}}+|p_{x=1}|\log{|p_{x=1}|}.
\end{eqnarray}
It is easy to see that $H(X)$ is always negative ($0$ for $\Delta=0$) and its asymptotics for large $\Delta$ is $-(\frac{1}{\Delta}+\log{\Delta})$. 

Nebit's entropy has some appealing intuition behind it. It is the minimal entropy you need to pump into your nebit to hide its negativity. Interestingly, this minimal entropy happens to be the entropy of the thermal distribution of a one dimensional string with temperature $T$ proportional to $\ln^{-1}{(\frac{1+\Delta}{\Delta})}$.

%%%%%%%%%%%%%%%%%%%%%%%%%%%%%%%%%%%%%%%%%%%%%%%%%%%%%%%%%%%%%%%%%%%%%%%%%%%%%%%%%%%%%%%%%%%%%%%%%%%%%%%%%%%%%%%%

%\section{More than one nebit and general negative probability distributions}

\subsection{Negativity catalysis}

In the following sections we are going to show that nebits can be used to improve efficiency of some classical protocols and that this efficiency depends on the amount of negativity $\Delta$. It is therefore natural to ask if it is possible to increase $\Delta$, say, with the help of another nebit.

Let us consider two independent nebits, the first one described by probabilities $\{1+\Delta,-\Delta\}$ and the second one by $\{1+\Delta',-\Delta'\}$. Their corresponding probability vectors are
\begin{equation}
\Pi_2 = \begin{pmatrix}
1 + \Delta \\ - \Delta
\end{pmatrix}, ~~\Pi'_2 = \begin{pmatrix}
1 + \Delta' \\ - \Delta'
\end{pmatrix},
\end{equation}
where the subscript '2' denotes a two-state system. The statistical independence implies that these two nebits can be jointly considered as a four-state system with the probability vector
\begin{equation}
\Pi_4 = \Pi_2 \otimes \Pi'_2 = \begin{pmatrix}
(1 + \Delta)(1 + \Delta') \\ - (1 + \Delta)\Delta' \\ - \Delta(1 + \Delta') \\ \Delta\Delta'
\end{pmatrix}.
\end{equation}

Next, consider a CNOT operation on both nebits
\begin{equation}
\begin{pmatrix}
(1 + \Delta)(1 + \Delta') \\ - (1 + \Delta)\Delta' \\ - \Delta(1 + \Delta') \\ \Delta\Delta'
\end{pmatrix} \rightarrow \begin{pmatrix}
(1 + \Delta)(1 + \Delta') \\ - (1 + \Delta)\Delta' \\ \Delta\Delta' \\ - \Delta(1 + \Delta')
\end{pmatrix}.
\end{equation}
After the CNOT the resulting marginal probability distributions are
\begin{equation}
\Pi_2 = \begin{pmatrix}
1 + \Delta \\ - \Delta
\end{pmatrix}, ~~\Pi'_2 = \begin{pmatrix}
1 + \Delta + \Delta' + 2\Delta\Delta' \\ - \Delta - \Delta' - 2\Delta\Delta'
\end{pmatrix}.
\end{equation}
We see that the control nebit does not change but the second nebit becomes more negative. This logical nebit operation can be interpreted as catalysis of negativity. Interestingly, one is unable to use the same two nebits in the second catalysis process because of correlations created by CNOT. The second application of the CNOT operation reverses the catalysis and restors the nebits to the initial state. The CNOT catalysis only works for independent nebits.

\subsection{Creation of general negative probability distributions}
We are also going to show that in some cases it is useful to work with signed probability distributions over more than two states. We are therefore going to present how to obtain such distributions from a single nebit.

Consider a distribution $\{p_1,\ldots,p_N\}$ such that
\begin{equation}
p_1 \geq p_2 \geq \ldots \geq p_k \geq p_{k+1} \geq \ldots \geq p_N,
\end{equation}
where $p_k \geq 0$ and $p_{k+1} < 0$. Moreover
\begin{equation}\label{negdistr}
\sum_{j=1}^k p_j = 1+\Delta,~~~~\sum_{j=k+1}^N p_j = -\Delta.
\end{equation}
Below we show how to generate this distribution starting from a nebit with a distribution $\{1+\Delta,-\Delta\}$.

Let us consider two stochastic processes: $S_1$ and $S_2$. The process $S_1$ starts with a single event that occurs with probability one and generates a probability distribution $\{q_1,\ldots,q_k\}$, where $q_j = \frac{p_j}{1+\Delta}$, distributed over the first $k$ events; $S_2$ starts with a single event and generates a probability distribution $\{r_1,\ldots,r_{N-k}\}$, where $r_j = -\frac{p_{k+j}}{\Delta}$, over the remaining $N-k$ events. To generate $\{p_1,\ldots,p_N\}$ one simply applies $S_1$ or $S_2$, depending on the flip of the nebit coin.

%%%%%%%%%%%%%%%%%%%%%%%%%%%%%%%%%%%%%%%%%%%%%%%%%%%%%%%%%%%%%%%%%%%%%%%%%%%%%%%%%%%%%%%%%%%%%%%%%%%%%%%%%%%%%%%%

\section{Classical dynamics with nebits}

Dynamics of a classical system can be represented as a trajectory in the state space. As we are interested in computational algorithms we limit ourselves to discrete state spaces and discrete time. This is not a serious limitation and one can translate our results to the continuous state space.

We enumerate time steps with integers, $t=0,1,2,\ldots$ and thus a $T$-step state space trajectory is a sequence of states
\begin{equation}
[s_0,s_1,\ldots,s_T],
\end{equation}
where $s_0$ is the initial state of the system. The first step takes the system from $s_0 \rightarrow s_1$ and so on. After $T$ steps the system ends up in the state $s_T$. 

The dynamics does not have to be reversible but we assume that it is deterministic, i.e., for any given state there exist a unique state to which the system is transformed to in the next step. Indeed, this is what defines a trajectory. For example, a trajectory corresponding to a reversible dynamics can look like this 
\begin{equation}
[a,b,c,d,\ldots],
\end{equation}
whereas a trajectory corresponding to an irreversible dynamics, given that the transition rules are time-independent, can look as
\begin{equation}
[a,b,b,b,\ldots ].
\end{equation}
The irreversibility in the latter follows from the loss of information in the state b about its predecessor: has it been a or b? 

Randomness in a deterministic dynamics is the result of observer's ignorance about the observed system. It can have two different origins. The first one is insufficient preparation and measurement readouts precision, in which case the initial, intermediate and final states are statistical ensembles. The second one is insufficient knowledge of the exact dynamics of the system such that at some point one is unsure if the correct transition rule is $a\rightarrow b$ or $a \rightarrow c$. In the first case, instead of a single trajectory one follows a collection of trajectories in the state space. In particular, one follows the evolution of a whole region of the state space. For reversible dynamics the size/cardinality of this region is conserved (Liouville theorem) but this may not be true for irreversible systems. In the second case one  observes splitting and merging of the trajectories such as in the Brownian motion, which can be modelled by a random walk. In this model the trajectories split because one is not sure whether some complex external forces make the particle move to the left or to the right. 

Before we introduce the aforementioned nebit black magic, let us go through a simple example illustrating the exact workings of the model considered in this work. This may be perceived by some readers as an unnecessary pedantism but one can never be too careful with non-orthodox ideas where intuitions have to be built up afresh. 

We start with the state space consisting of $N$ states ${\mathcal S} = \{a_1,a_2,\ldots,a_N \}$ and choose the following cyclic and reversible transition rule $a_i \rightarrow a_{i+1}$ with $a_N \rightarrow a_1$. If the system is initialised in the state $a_7$ and it evolves for a sufficiently long time its trajectory is
\begin{equation}
[a_7,a_8,\ldots,a_N,a_1,\ldots,a_6,a_7,a_8,\ldots].
\end{equation} 

Now, if the initial state of the system is not precisely defined, say, it is $a_3$ with probability $p$ and $a_7$ with probability $1-p$ one ends up with a mixture of trajectories
\begin{equation}
p[a_3,a_4,a_5,\ldots] + (1-p)[a_7,a_8,a_9,\ldots].
\end{equation}
Therefore, after two steps the system is in the state $a_5$ with probability $p$ and $a_9$ with probability $1-p$.

Next, let us consider a different situation. Imagine that the system is initiated in the above random state and that due to some external influences there is a probability $1/2$ that in one step the transition rule will be applied twice. In other words, before each step one tosses a fair coin and if the result is heads one applies the transition rule once, but if its tails one applies the transition rule twice. Therefore, after one step the mixture of the trajectories is
\begin{equation}
\frac{p}{2}[a_3,a_4] + \frac{p}{2}[a_3,a_5]+\frac{1-p}{2}[a_7,a_8] + \frac{1-p}{2}[a_7,a_9],
\end{equation} 
after two steps it is
\begin{eqnarray}
& &\frac{p}{4}[a_3,a_4,a_5] + \frac{p}{4}[a_3,a_5,a_6]+\frac{p}{4}[a_3,a_4,a_6] +  \nonumber \\
& &\frac{p}{4}[a_3,a_5,a_7]+\frac{1-p}{4}[a_7,a_8,a_9] + \frac{1-p}{4}[a_7,a_9,a_{10}] + \nonumber \\
& &\frac{1-p}{4}[a_7,a_8,a_{10}] + \frac{1-p}{4}[a_7,a_9,a_{11}],
\end{eqnarray} 
and so on. After two steps the probability of finding the system in the state $a_6$ is $\frac{p}{4} + \frac{p}{4} = \frac{p}{2}$, since there are two trajectories going to this state and we need to sum up their corresponding probabilities.

%%%%%%%%%%%%%%%%%%%%%%%%%%%%%%%%%%%%%%%%%%%%%%%%%%%

\subsection*{Trajectories with negative probabilities}

Now we inject negative probabilities into the dynamics. We start with the following observation: in the previous example there was an obvious assumption that we did not trace trajectories whose probabilities were zero. Such trajectories simply did not occur. However, this assumption is not that obvious when negative probabilities come into play -- one needs to be very careful what is traced and what is not.

In order to get some intuition about what can happen let us start with a simple example that leads to troublesome interpretations, which we would like to avoid in the future. Imagine a single step of a random evolution that allows for trajectories with negative probabilities. The system can follow one of four possible trajectories, first three with probability $1/2$ and the last one with probability $-1/2$
\begin{equation}
\frac{1}{2}\left([a_0,a_0]+[a_0,a_1]+[a_{2},a_{2}]-[a_{2},a_{3}] \right).
\end{equation}
At $t=0$ the system is in $a_0$, since the probability of finding it in this state is $\frac{1}{2}+\frac{1}{2}=1$. There are also trajectories starting from $a_{2}$, but we do not observe this state, since the corresponding probability is $\frac{1}{2}-\frac{1}{2}=0$. Therefore, in the orthodox scenario with only positive probabilities these trajectories would not be considered. However, at $t=1$ the trajectories split and we suddenly observe that the system can be in one of the four different states: $a_0$, $a_1$ and $a_{2}$, each with probability $1/2$, and $a_{3}$ with probability $-1/2$. From the point of view of an observer a single trajectory starting from $a_0$ splits into two and another two trajectories spontaneously emerge from $a_{2}$. The problem we do not want to deal with is how to interpret trajectories with negative probabilities as well as events with inflated probabilities -- e.g. events that are complementary to events with negative probabilities.  

In order to avoid these problems we need to set some restrictions on possible dynamics so that negative and inflated probabilities are never directly observed. Perhaps one of the simplest solutions is to impose that whenever a negative probability trajectory and a positive probability trajectory split, the negative probability trajectory must immediately merge with some other positive probability trajectory. This way negative probabilities will always be 'hidden under' positive probabilities. Therefore, in our approach we allow for some spontaneous emergence of negative trajectories, provided that they are always properly compensated with the positive ones. We study in more details some examples of such dynamics in the next sections.

%%%%%%%%%%%%%%%%%%%%%%%%%%%%%%%%%%%%%%%%%%%%%%%%%%%%%%%%%%%%%%%%%%%%%%%%%%%%%%%%%%%%%%%%%%%%%%%%%%%%%%%%%%%%%%%%

\section{Negative random walks}

Let us consider $T$ steps of a dynamics generated by a random distribution over some set of $K$ trajectories 
\begin{equation}
\sum_{j=1}^K p_j [a_{0}^{(j)},a_{1}^{(j)},\ldots,a_{T}^{(j)}],
\end{equation}
where $a_{m}^{(j)} \in {\mathcal S} = \{a_1,a_2,\ldots,a_N \}$ corresponds to the state of the system after m-th step along the j-th trajectory. This trajectory occurs with probability $p_j$, which can be negative. Let us assume that the distribution $\{p_1,\ldots,p_K\}$ is analogous to the distribution described by Eq. (\ref{negdistr}) which can be generated with the help of a single nebit. 

The above evolution can be interpreted as a kind of a random walk on the system's state space ${\mathcal S}$. We call it a {\it negative random walk}. In order to be sure that one never observes negative and inflated probabilities we need to set some restriction on trajectories and the distribution $\{p_1,\ldots,p_K\}$. To do this let us note that the probability that after $m$ steps the system is in the state $a_k$ is 
\begin{equation}
p(a_k|m)=\sum_{j=1}^{K}p_j \gamma_{jmk},
\end{equation}
where $\gamma_{jmk} = 1$ if $a_{m}^{(j)}=a_k$ and $\gamma_{jmk} = 0$ if $a_{m}^{(j)}\neq a_k$. Therefore, the restriction takes the following form
\begin{equation}\label{restriction}
\forall_{0 \leq m \leq T} ~~p(a_k|m) \geq 0.
\end{equation}

%%%%%%%%%%%%%%%%%%%%%%%%%%%%%%%%%%%%%%%%%%%%%%%%%%%

\subsection*{Super-ballistic negative random walk on a chain}

A single step of the above random walk can in principle take a system from any state to any other state. However, for many physically motivated examples the topology of a state space ${\mathcal S}$ is not arbitrary and is given by some graph. This graph determines which states can be placed on subsequent positions for any allowed trajectory. 

Perhaps the most studied example of such a graph is a chain graph, i.e., a segment of a one-dimensional discrete space in which positions are described by integers $x=1,2,\ldots, N$. The chain graph is often used to study diffusion in one-dimensional space. The system can be interpreted as a particle walking on a discrete line where the states represent positions $a_k \equiv k$. In a single step the particle can change its position from $x$ to $x\pm 1$ or stay in the same place (with the exception of boundaries -- if $x=1$ the next position can be either 1 or 2 and if $x=N$ the next position can be either $N$ or $N-1$). 

Let us first consider a standard random walk in which the particle is initially localized at some position. If we prepare a uniform probability distribution over all trajectories starting from the initial position we will observe a diffusive spread. However, we can manipulate this probability distribution to obtain a ballistic spread. Diffusive spread means that the standard deviation of the spatial probability distribution is proportional to the square root of the number of steps, whereas ballistic means that it is proportional to the number of steps. Due to the topology of a chain graph a random walk on it can be at most ballistic. The greatest spread can be achieved by the following random walk. Consider $T=N-1$ steps of a walk generated by an even mixture of two trajectories
\begin{equation}\label{crw}
\frac{1}{2}[1,1,1,\ldots,1,1] + \frac{1}{2}[1,2,3,\ldots,N-1,N].
\end{equation}
The particle starts at position $x=1$ and then it either stays at $x=1$ or always moves one step to the right. Each possibility occurs with probability $1/2$. After $T$ steps the particle is in an even mixture of being at $x=1$ and $x=N$. It is straightforward to show that the standard deviation of the position after $m$ steps is
\begin{equation}\label{sigmap}
\sigma (m) = \frac{m}{2}.
\end{equation}

\begin{figure}[t]
\includegraphics[scale=0.60]{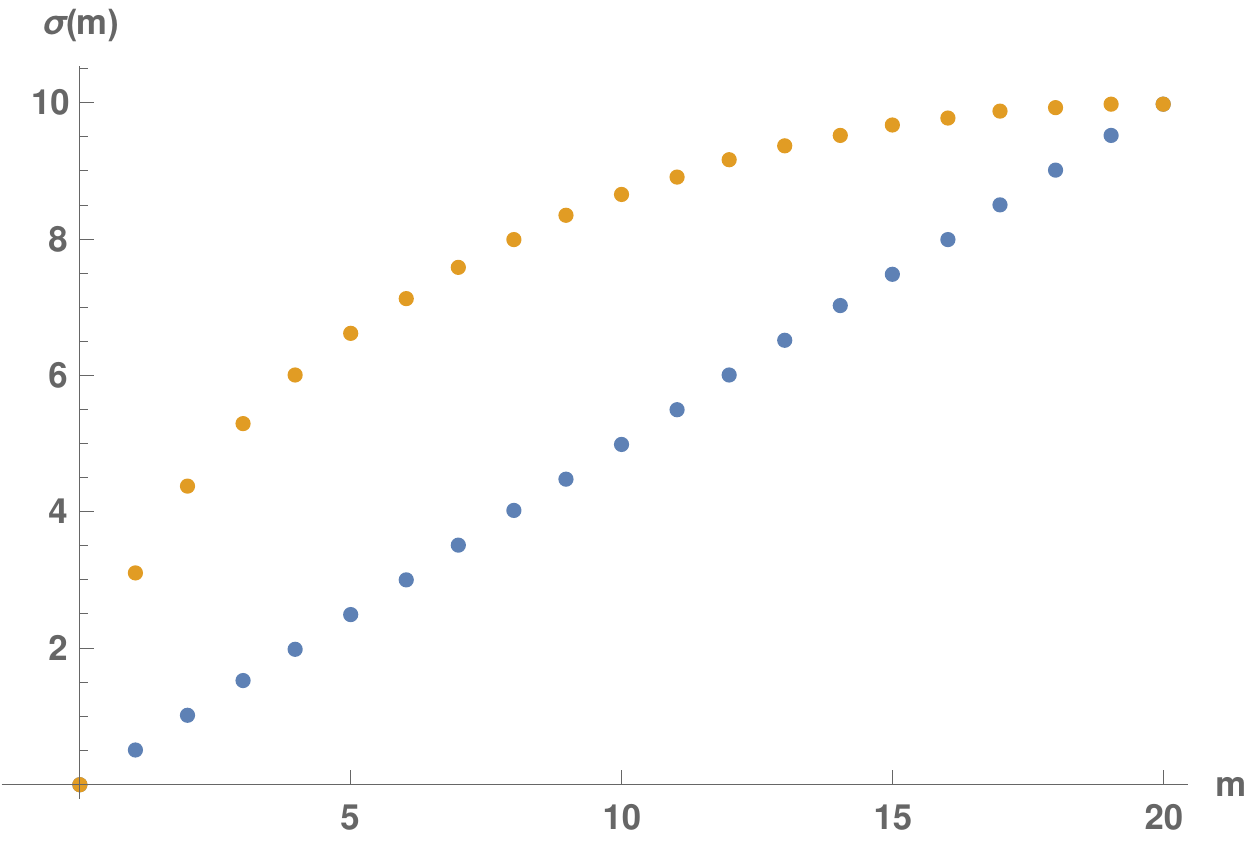}
\caption{Standard deviation of the position for $T=20$ steps of a walk on a chain of length $N=21$. Orange dots correspond to a negative random walk and blue dots correspond to a classical random walk described by Eq (\ref{crw}). Negative random walk exhibits super-ballistic behaviour. \label{fig1}}
\end{figure}

Next we show the corresponding negative random walk on a chain can spread much faster, i.e., it can be super ballistic. This effect stems from the spontaneous emergence of trajectories that we observed in the previous section, however this time we are going to show how to avoid direct observation of negative probabilities. We consider $T=N-1$ steps along the following mixture of trajectories
\begin{eqnarray}
& & [1,\ldots,1]+\frac{1}{2}[N,\ldots,N] + \delta \sum_{x=2}^{N-1} [x,\ldots,x] \nonumber \\
& & - \delta \sum_{x=2}^{N-1} [\underbrace{x,x-1,\ldots,2,1}_{x},\underbrace{1,\ldots,1}_{T+1-x}] \nonumber \\
& & - \delta \sum_{j=0}^{T-1}[\underbrace{N,\ldots,N}_{j},\underbrace{N,N-1,\ldots,N-T+j}_{T+1-j}].
\end{eqnarray}
The trajectories that occur with positive probabilities represent a particle that stays at a fixed position. The only movement is generated by trajectories that occur with negative probabilities. The sum of all positive probabilities in the above distribution is $\frac{3}{2} + (N-2)\delta$, whereas the sum of all negative ones is $-(N-2)\delta -T\delta$. Since all probabilities need to add up to one we conclude that $T\delta = \frac{1}{2}$, hence $\delta = \frac{1}{2T}$. The probability that after $m$-th step the particle is at position $x$ is given by
\begin{eqnarray}
p(x|m) &=& 0 ~~ \text{for} ~~ 1<x<N, \\
p(1|m) &=& 1 - \frac{t}{2T}, \\
p(N|m) &=& \frac{t}{2T}.
\end{eqnarray}
This is a quite counter-intuitive nonlocal process. The particle does not move through the chain but rather jumps directly from $x=1$ to $x=N$, seemingly ignoring the topology of the graph. Interestingly, at the beginning and at the end of the walk the above probabilities are the same as in the standard positive probability case considered above. However, it is straightforward to show that in this case the standard deviation of the position after $m$ steps is
\begin{equation}\label{sigman}
\sigma (m) = \frac{1}{2}\sqrt{m(2T-m)}.
\end{equation}    
In Fig. \ref{fig1} we plot standard deviations (\ref{sigmap}) and (\ref{sigman}) and show that negative random walk on a chain is super-ballistic. 

Finally, we should mention at this point the research on quantum walks. Quantum walks are quantum counterparts of classical random walks that take advantage of the interference phenomenon and are known to spread faster than classical ones (though still ballistically on a chain graph). The fast spreading can be used to construct efficient quantum algorithms \cite{QW1,QW2}. We speculate that many features of quantum walks can be simulated with classical random walks supplemented with nebits.

%%%%%%%%%%%%%%%%%%%%%%%%%%%%%%%%%%%%%%%%%%%%%%%%%%%%%%%%%%%%%%%%%%%%%%%%%%%%%%%%%%%%%%%%%%%%%%%%%%%%%%%%%%%%%%%%

\section{Search with negative probabilities}

Consider a database consisting of $N$ elements. We assume that these elements correspond to states of some system ${\mathcal S} = \{a_1,a_2,\ldots,a_N \}$. Next imagine that there is a marked element, say the state $a_N$, and our goal is to find it. 

Let us first discuss a standard probabilistic search algorithm. In this case the optimal search method is the simplest one. We are going to evolve the system through all states and check whether the state of the system is marked or not. Of course, the number of elements we need to check before we find the marked one depends on the way we order them, but since we do not know in advance which element is marked, the best solution is to consider a random order. Therefore, we prepare an even mixture of all trajectories
\begin{equation}\label{searchinit}
\frac{1}{N!}\sum_{\pi}[a_{\pi(1)},a_{\pi(2)},\ldots,a_{\pi(N)}],
\end{equation}
and evolve the system. In the above the sum is taken over all $N!$ permutations of states $\pi(i)$.  A single step of the protocol consists of two parts. First, we check the state of the system. If it is in the marked state $a_N$ we stop the protocol and announce: {\it FOUND}. If it is not, we {\it PROCEED} to the next state along the trajectory. 

What is the probability that the announcement is made after exactly $t$ steps of the evolution? There are $(N-1)!$ trajectories for which $a_N$ is at position $t$, therefore the probability that the announcement is made exactly after $t$ steps is $1/N$. Hence, the average search time is
\begin{equation}
\bar{T}=\sum_{t=1}^N \frac{t}{N} = \frac{N+1}{2}.
\end{equation}
Moreover, the probability that the announcement is made after $t$ steps, or earlier, is $t/N$. Therefore, it takes $N$ steps to be sure that the marked state is found. 

Next, let us consider a strategy we could use if we had access to a nebit. Let us start with the same initial mixture of trajectories (\ref{searchinit}) as before. The new strategy is a simple modification of the previous one. This time if we find that the system is in the marked state we check the outcome of the nebit. If the outcome is zero (with probability $1+\Delta$) we announce: {\it FOUND}. However, if the outcome is one (with probability $-\Delta$) we {\it PROCEED} to evolve the system along the trajectory. 

First, note that the probability that the announcement is made exactly after $t$ steps is $(1+\Delta)/N$. This is due to the same reason as before and the $\Delta/N$ improvement comes from the use of the nebit. The probability that the announcement is made after $t$ steps, or earlier, is $t(1+\Delta)/N$. This time it takes $\tau =\lfloor N/(1+\Delta) \rfloor$ steps to be almost sure that the marked state is found. The average search time is
\begin{equation}
\bar{T}=\sum_{t=1}^{\tau} \frac{t(1+\Delta)}{N} = \frac{\tau+1}{2}.
\end{equation}
Clearly, for $\Delta = \sqrt{N}$ we achieve the efficiency of the Grover search algorithm, but in principle we can achieve our goal in a single step if $\Delta = N-1$.

It is obvious that the achieved speedup is due to the inflated probability $(1+\Delta)$. What is not obvious is that the proposed protocol, that runs for $\tau$ steps, does not allow one to directly observe negative nor inflated probabilities. To prove it, note that because of the symmetry the probability that the system is in an unmarked state is the same for all unmarked states. Say we focus on a state $a_1$. We already known that there are $(N-1)!$ trajectories for which $a_1$ is at position $t$. And because we have used the nebit we need to count for how many of such trajectories the marked state $a_N$ precedes $a_1$. If we fix positions of $a_1$ and $a_N$ we get $(N-2)!$ different trajectories. Given that $a_1$ is at position $t$, $a_N$ can take one of the $t-1$ preceding positions or one of the $N-t$ remaining positions. A trajectory for which $a_N$ follows $a_1$ occurs with the probability $\frac{1}{N!}$, but a trajectory for which $a_N$ precedes $a_1$ occurs with the probability $-\frac{\Delta}{N!}$. Therefore, the probability that during the $t$-th step the system is in the state $a_1$ is
\begin{equation}
p(t)= \frac{N-t-\Delta(t-1)}{N(N-1)}.
\end{equation}  
This probability decreases linearly in $t$ and it reaches zero for $t=\frac{N+\Delta}{1+\Delta} \geq \tau$. 

Achieving the un-realistic single step speedup comes at the cost of the negative nebit entropy of around $-\frac{\log{N}}{N}$ per the database size for a large $N$. The Grover speedup cost is half of that: $(-\frac{\log{N}}{N})\times 0.5$. This is of course in line with our intuition about the absence of free lunch - the more speedup you need the more nebits you have to inject into the system although the total amount of negativity pumped in per 'volume' of the database tends to zero with $N\rightarrow \infty$.

%%%%%%%%%%%%%%%%%%%%%%%%%%%%%%%%%%%%%%%%%%%%%%%%%%%%%%%%%%%%%%%%%%%%%%%%%%%%%%%%%%%%%%%%%%%%%%%%%%%%%%%%%%%%%%%%

\section{Discussion}
In this paper we follow a well established tradition in theoretical physics of playing with non-orthodox ideas to gain some insights into the behaviour of complicated physical systems. Our main purpose is to find a simple classical simulation of some aspects of quantum computation with the help of a hypothetical nebit - a binary system with signed probabilities. We are not interested in philosophy of nebits but only in their formal mathematical properties necessary to achieve quantum speedup using simple, classical dynamical systems. Nebits are a mathematical tool and nothing more.

We hope that the insights we can gain from studying nebits will show us different ways of understanding quantum computing and quantum mechanical processes in general. What immediately springs to mind is (a) quantitative classification of the quantumness of quantum computing algorithms in relation to their classical counterparts and (b) a way to generate new quantum algorithms from nebit supplemented classical ones. We already elaborated on (a)
 in this manuscript calculating the nebit cost of the Grover search algorithm (analysis of quantum contextuality and other quantum algorithms in preparation). It will be interesting to elaborate on (b) but this is the scope of our future research.

%%%%%%%%%%%%%%%%%%%%%%%%%%%%%%%%%%%%%%%%%%%%%%%%%%%%%%%%%%%%%%%%%%%%%%%%%%%%%%%%%%%%%%%%%%%%%%%%%%%%%%%%%%%%%%%%

\section*{Acknowledgements}

DK is supported by National Research Foundation, Prime Ministers Office, Singapore and the Ministry of Education, Singapore under the Research Centres of Excellence programme. PK was supported by the Ministry of Science and Higher Education in Poland (science funding scheme 2016-2017 project no. 0415/IP3/2016/74). 

%%%%%%%%%%%%%%%%%%%%%%%%%%%%%%%%%%%%%%%%%%%%%%%%%%%%%%%%%%%%%%%%%%%%%%%%%%%%%%%%%%%%%%%%%%%%%%%%%%%%%%%%%%%%%%%%

%%%%%%%%%%%%%%%%%%%%%%%%%%%%%%%%%%%%%%%%%%%%%%%%%%%%%%%%%%%%%%%%%%%%%%%%%%%%%%%%%%%%%%%%%%%%%%%%%%%%%%%%%%%%%%%%
%%%%%%%%%%%%%%%%%%%%%%%%%%%%%%%%%%%%%%%%%%%%%%%%%%%%%%%%%%%%%%%%%%%%%%%%%%%%%%%%%%%%%%%%%%%%%%%%%%%%%%%%%%%%%%%%


\begin{thebibliography}{99}
\bibitem{Shor} P. W. Shor, Proceedings of the 35th Annual Symposium on Foundations of Computer Science, Santa Fe, NM, Nov. 20--22, (1994).
\bibitem{Grover} L. K. Grover., Proceedings of the twenty-eighth annual ACM symposium on Theory of Computing, 212-219 (1996). 
\bibitem{bosonsampling} S. Aaronson and A. Arkhipov, Theory of Computing {\bf 9}, 143 (2013).
\bibitem{Benioff} P. Benioff, J. Stat. Phys. {\bf 22}, 563 (1980).
\bibitem{Google} F. Arute {\it et. al.}, Nature {\bf 574}, 505 (2019).
\bibitem{Szekely1} I. Z. Ruzsa and G. J. Székely, Algebraic Probability Theory, Wiley, New York (1988).
\bibitem{Szekely2} G. J. Székely, Half of a coin: negative probabilities. Wilmott Magazine, 66–68, (2005).
\bibitem{reviewonnegative} W. M{\"u}ckenheim {\it et. al.}, Physics Reports (Review Section of Physics Letters) 133, {\bf 6} (1986).

\bibitem{Dirac}  P.A.M. Dirac, {\it The physical interpretation of quantum mechanics} Proceedings of the Royal Society of London Series A, Mathematical and Physical Sciences, {\bf 180}, 1 (1942).

\bibitem{Feynman} R. P. Feynman, J. Theor. Phys. {\bf 21}, 467 (1982).

\bibitem{Wigner} E. Wigner, Phys. Rev. {\bf 40},  749 (1932).

\bibitem{neg1} A. Khrennikov, On negative probabilities, in {\it Derivatives: Models on Models} Wiley and Sons, Ltd., Chichester, 317-323 (2007)

\bibitem{neg2} M. Burgin. Interpretations of negative probabilities. arXiv preprint 1008.1287, (2010).

\bibitem{neg3} S. Abramsky and A. Brandenburger, The sheaf-theoretic structure of non-locality and contextuality, New Journal of Physics, {\bf 13}, 113036 (2011).

\bibitem{neg4} S. Abramsky, A. Brandenburger, in \textit{Horizons of the Mind: A Tribute to Prakash Panagaden}, ed. F. van Breugel and E. Kashefi and C. Palamidessi and J. Rutten, Springer, pages 59--75, (2014).

%\bibitem{neg5} J. A. de Barros, Decision Making for Inconsistent Expert Judgments
%Using Negative Probabilities, Lecture Notes in Computer Science,
%257–269. Springer, Berlin/Heidelberg, (2014).

%\bibitem{neg6}  J. A. de Barros and G. Oas, Negative probabilities and counterfactual
%reasoning in quantum cognition, Physica Scripta, {\bf 163}, 014008 (2014).

\bibitem{neg7}  G. Oas, J. A. de Barros, and C. Carvalhaes, Exploring non-signalling
polytopes with negative probability. Physica Scripta, {\bf 163}, 014034 (2014).

%\bibitem{neg8}  J. A. de Barros and G. Oas, Quantum Cognition, Neural Oscillators,
%and Negative Probabilities, in Emmanuel Haven and Andrei Khrennikov,
%editors, The Palgrave Handbook of quantum models in social science: applications
%and grand challenges. Palgrave MacMillan, (2015).

\bibitem{neg9} J. A. de Barros, G. Oas, and P. Suppes, Negative probabilities
and Counterfactual Reasoning on the double-slit Experiment. In J.-Y. Beziau, D. Krause, and J.B. Arenhart, editors, Conceptual Clarification:Tributes to Patrick Suppes (1992-2014). College Publications, London, (2015). 

\bibitem{neg10} J. A. de Barros, G. Oas, Negative probabilities and contextuality, J. Math. Psych. {\bf 74}, 34, (2016).

\bibitem{QW1} A. Ambainis, Int. J. Quant. Inf. {\bf 1}, 507 (2003). 
\bibitem{QW2} M. Santha, 5th Theory and Applications of Models of Computation (TAMC08), Xian, April 2008, LNCS 4978, 31 (2008).
\end{thebibliography}
\end{document}